\documentstyle[12pt,aaspp4,flushrt]{article}
\def\etal{{\it et~al.~}}
\def\eg{{\it e.g.~}}

\begin{document}

\title{{\it Hubble Space Telescope} Observations of Globular Clusters
in M31 II: Structural Parameters\footnote[1]{Based on observations
with the NASA/ESA Hubble Space Telescope, obtained at the Space
Telescope Science Institute, which is operated by AURA, Inc., under
NASA contract NAS 5-26555.}}

\author{Carl J. Grillmair\altaffilmark{2}}
\affil{UCO/Lick Observatory, University of California, Santa Cruz, CA 95064}

\altaffiltext{2}{Current Address: Jet Propulsion Laboratory,
California Institute of Technology, Mail Stop 183-900, 4800 Oak Grove
Drive, Pasadena, CA 91109-8099}

\author{Edward A. Ajhar}
\affil{KPNO, National Optical Astronomy Observatories
\footnote[3]{Operated by AURA under cooperative agreement with the
National Science Foundation.}, P.O.Box 26732, Tucson, AZ 85726}
\author{Sandra M. Faber}
\affil{UCO/Lick Observatory, Board of Studies in Astronomy and
Astrophysics, University of California, Santa Cruz, CA 95064}
\author{William A. Baum}
\affil{University of Washington, Astronomy Department, Box 351580,
Seattle, WA 98195}
\author{Jon A. Holtzman}
\affil{Astronomy Department, New Mexico State University, Box 30001 /
Dept. 4500, Las Cruces, NM 88003}
\author{Tod R. Lauer, C. Roger Lynds, and Earl J. O'Neil, Jr.}
\affil{KPNO, National Optical Astronomy Observatories$^3$,
P.O.Box 26732, Tucson, AZ 85726}

\begin{abstract}

	We analyze post-refurbishment Hubble Space Telescope images of
four globular clusters in M31. The ability to resolve stars to below
the horizontal branch permits us to use star counts to extend the
surface brightness profiles determined using aperture photometry to
almost 5 orders of magnitude below the central surface density. Three
of the resulting cluster profiles are reasonably well-fit using
single-mass King models, with core and tidal radii typical of those
seen in Galactic globular clusters. We confirm an earlier report of
the discovery of a cluster which has apparently undergone core
collapse.  Three of the four clusters show departures in their
outskirts from King model behavior which, based on recent results for
Galactic globulars, may indicate the presence of tidal tails.

\end{abstract}


\section{Introduction.}

	Globular clusters have been and continue to be useful tools in
the study of the structure, formation, dynamics, and distances of
galaxies. The clusters themselves are dynamically simple systems, and
are sufficiently luminous to be visible out to $cz \approx$ 10000
km/sec (Harris 1991). Much recent work has concentrated on the degree
to which the properties of globular clusters differ from one galaxy to
the next. For example, an important issue currently being addressed is
whether the globular cluster luminosity function and its peak are
sensitive to metallicity (\eg Ashman \etal 1995), and observably
different for early- and late-type galaxies. Considerable work is
being done to better understand the physical processes which would
alter the structure and dynamics of individual clusters, and thereby
the global properties of globular cluster {\it systems}. For example,
different orbit shapes, tidal fields, and the presence of disks and
bulges may have profound consequences for the ultimate survival of
individual clusters, and for the constitution of cluster systems as a
whole (Chernoff \& Weinberg 1990; Kundic \& Ostriker 1995). While
hypotheses are in abundance, the observations necessary to test these
ideas remain rather sparse. We need to make detailed studies of
globular clusters in differing environments if we are to put useful
constraints on the models.

	The surface brightness profiles of most globular clusters in
our own Galaxy have long been known to be reasonably well represented
by King models (King 1966), which require the minimum possible number
of parameters to describe the central density, total energy, and
external tidal field. Together, the measured core or half-light radii
and central densities yield information concerning the formation and
long-term dynamical evolution of individual clusters. Conversely, the
limiting radii are sensitive, on a relatively short time scale, to the
influence of external potentials. The actual form of the observed
tidal ``cutoff'' is determined by cluster mass, the tidal field of the
galaxy, and the shape and phase of the cluster orbit (Oh and Lin 1992;
Grillmair \etal 1996). While these quantities are not well constrained
for any one cluster, obtaining information on a large sample of
clusters should enable us to describe the mean orbital properties of
the ensemble, as well as to map the potential of the host galaxy. M31
globular clusters are especially interesting in this respect because
$(i)$ the M31 cluster system is the most populous cluster system in
the Local Group, $(ii)$ we observe the cluster system from a vantage
point outside the system, and $(iii)$ M31 is similar in many respects
to our own Galaxy, permitting us to gauge the sensitivity of globular
cluster properties to detailed structural differences.

	The superb resolution afforded by the Hubble Space Telescope
(HST) since the installation of the Wide Field/Planetary Camera-2
(WFPC-2) permits study of the stellar content and distribution in
extragalactic globular clusters at a level of detail hitherto only
possible for clusters belonging to our own Galaxy and its
satellites. The observations discussed below were motivated primarily
by a desire to calibrate the metallicity dependence of horizontal
branch (HB) magnitudes (Ajhar \etal 1994; Ajhar \etal 1996, hereafter
Paper I). However, the spatial resolution and the depth of the
observations also allow a detailed examination of the structure of the
clusters themselves. The mass/luminosity ratios derived from velocity
dispersion measurements of our sample clusters are discussed elsewhere
(Dubath and Grillmair 1996).

	Section 2 briefly summarizes the observations. Section 3
describes synthetic aperture photometry, and a star-count analysis is
carried out in Section 4. We present the results and discuss their
implications in Section 5, and briefly summarize our findings in
Section 6.

\section{Observations. \label{sec:obs} }

	As one of the primary motivations for HST imaging of globular
clusters in M31 was to calibrate the metallicity dependence of
horizontal branch (HB) magnitudes, the clusters we chose to observe
were selected to give a wide range in metallicity and to be as clear
of the disk of M31 as possible. The four clusters we examined are
discussed in detail in Paper I, and we simply summarize the basic data
concerning our targets in Table 1. Cluster designations refer to the
list of Sargent \etal (1977) while magnitudes and metallicities are
from Huchra \etal (1991).

	The images we discuss were taken in February, 1994, shortly
after the WFPC-2 was installed in the HST.  Two 1000s exposures in each
of F555W ($V$) and F814W ($I$) were taken of each cluster. The
telescope was pointed so as to put the target clusters in the
Planetary Camera (PC), giving $0.0455\arcsec$ per pixel (Holtzman
\etal 1995a). The images were taken prior to the reduction in CCD
operating temperature from $-77^\circ$C to $-88^\circ$C, and accordingly
suffer from a somewhat larger number of hot pixels (that is, pixels
with abnormally high dark current) than is
presently the case. The majority of these were removed using maps of
the hot pixels developed by the WFPC-2 team. As discussed in Paper I,
the higher operating temperature also affects photometry, so a 0.05
magnitude offset was applied to the F814W zeropoint given in Holtzman
\etal (1995b).

\section{Aperture Photometry}

	Synthetic, concentric aperture photometry was carried out on
both the F555W and F814W images.  The centers of the clusters were
determined by computing the center-of-light in circular annuli of
varying radii (from 0.25-1.0\arcsec), and by subtracting fitted,
2-dimensional King models from the images. Aside from a few $\sim 1$
pixel excursions due to bright giants, the coordinates of center were
found to be reasonably uniform with radius, and are accurate
to within a pixel. Sky-background levels were estimated by computing
the mean counts in regions lying as far as possible from the clusters.
Owing to the large apparent size of the clusters in the PC, variations
in the distribution of background sources, the presence of bright
stars, and gradients due to the distribution of halo stars of M31, the
uncertainty in sky-background levels is about 5\%.  Aperture
photometry was carried out using a fixed center and circular
apertures. The uncertainties were computed based on the spread in
surface brightnesses measured in 8 sectors spread evenly around each
annulus, combined in quadrature with the estimated uncertainty in the
sky background. For the central pixel, the uncertainty was computed
based on both photon statistics and centering errors. Our surface
brightness measurements, transformed to $V$ and $I$ magnitudes using
the coefficients of Holtzman \etal (1995b), corrected for reddening
using the reddening law of DaCosta \& Armandroff (1990),

\begin{equation}
A_V : A_I : E_{B-V} = 3.200 : 1.858 : 1.000,
\end{equation}

\noindent and the values of $E_{B-V}$ given in Paper I, are tabulated
in Tables 2 through 5. We include only those radii for which the
estimated uncertainties are less than 0.25 mag.
	
	In Figure \ref{fig:color} we show the background-subtracted,
$V-I$ color profiles of the four clusters. No significant color
gradient is evident in any of the clusters, and the $\simeq 0.1$ mag
bluing apparent at the smallest radii in K58 and K105 is consistent
with scatter in the color distribution owing to the discrete
distribution of the brightest (and reddest) stars in the
cluster. However, this does not preclude the existence of a
substantial number of blue stragglers in the cores of these clusters
(which happen to be the two most concentrated in our sample).

	Careful inspection of the images of K108 and division of the
images by 2-dimensional King-model fits reveal what at first glance
appears to be a very narrow dust lane. The lane crosses the cluster
only $0.1\arcsec$ from the center, extending $\simeq 1.5\arcsec$ in
either direction.  In some respects it resembles the rings of dust
seen in HST images of the centers of elliptical galaxies (Grillmair
\etal 1994; van Dokkum \& Franx 1996)). On the other hand, K108 is
projected only $22\arcmin$ from the nucleus of M31, and a dust lane
might better be attributed to the intervening ISM of M31's disk.
Division of F555W by F814W reveals that the color of the feature is
actually slightly {\it bluer} than the surrounding starlight. This
would suggest that, rather than caused by obscuration, the dimming is
simply a consequence of a statistical dearth of red giant stars
over a small region.

\section{Star Counts}

	Owing to the low sky brightness in the WFPC-2 frames, the
aperture photometry extends reliably for some considerable distance
beyond the core radii of our clusters. However, at large radii the
uncertainties become dominated by the distribution of relatively few,
luminous giants. We consequently used star counts to enable us to
extend the radial profiles to even lower surface densities.

	Star counts were carried out using DAOPHOT II and ALLSTAR
(Stetson 1987; 1994a). Initial experiments were conducted using the
newer and more accurate ALLFRAME routine (Stetson 1994b), but owing to
$(i)$ the significantly increased amount of CPU time required to
obtain a list of magnitudes and colors, $(ii)$ the need for very
extensive completeness tests (see below), and $(iii)$ the similarity
in the color-magnitude distributions of cluster and field stars (which
ruled out separating cluster and field stars on the basis of color),
we concluded that DAOPHOT/ALLSTAR were of themselves sufficient for
our purposes.

	   Based on the sizes of globular clusters in the Milky Way,
we expected that the tidal radii of our sample clusters would be of
the same order as the field-of-view of the PC. Combined with the
off-center locations of the clusters in the PC and the presence of a
gradient in the surface density of field stars due to the halo of M31,
we elected to include all four WFPC-2 chips in our analysis. Objects
were detected and measured in the summed F555W and F814W images and
the output detection lists were used to compute magnitudes and colors
using the WFPC-2 zeropoints and color terms of Holtzman \etal
(1995b). Examination of images from which the measured stars had been
subtracted revealed that residual cosmic rays, visibly warm pixels,
and extended objects had, in the vast majority of cases, been so
identified and subsequently ignored by the software. We estimate that
less than 1\% of objects classified by DAOPHOT as stars could with
some degree of confidence be called galaxies, cosmic rays, or hot
pixels. Using the point spread function (PSF) index described by Baum
\etal (1995) we find that, at our ultimately selected magnitude cutoff
of $V$ = 25.5, we are well clear of the regime where spurious
detections become significant.  Only stars having both an F555W and an
F814W measurement and colors and magnitudes reasonably consistent with
those of cluster stars were counted. However, since the background is
overwhelmingly dominated by stars in the halo of M31 (whose
distribution over apparent magnitude and color is almost identical to
that of the cluster stars), little differentiation is possible.

	The detection and measurement of individual stars becomes
progressively more difficult as one approaches the crowded, inner
regions of each cluster.  For the PC images, magnitudes were
consequently measured using the crowded-field, PSF-fitting code
ALLSTAR. On the other hand, owing to the relatively uncrowded nature
of the outer fields, combined with the higher degree of undersampling
(making PSF-fitting problematic), magnitudes of sources detected in
the WF chips were measured using simple aperture
photometry. Naturally, the use of two different measuring algorithms,
combined with spatially varying noise and crowding levels, differing
pixel-scales, and zero-point offsets between chips, meant
that extensive simulations were required to combine the various
star-count results and surface brightness measurements reliably.

	The probability of detecting a particular star varies with
radius, magnitude, color, and chip number.  Completeness tests were
carried out by adding appropriately scaled PSFs to the images and
running the DAOPHOT II/ALLSTAR detection and measuring routines in a
manner identical to that applied to the real data.  We divided the
color-magnitude diagrams into a $1 \times 1$ magnitude grid as
shown in Figure \ref{fig:cm}. Experiments were conducted independently
for each cluster, each chip, and each color-magnitude bin, and
typically 200 artificial stars were added in the course of each
experiment. Artificial stars with a specified $V$ magnitude were added
to the F555W frame, and a matching set of stars with identical,
cluster-centric coordinates and $I$-magnitudes appropriate to the
specified color were added to the F814W frame. The DAOPHOT II/ALLSTAR
sequence was applied to both frames, and the output photometry files
were compared to determine how many artificial stars were retrieved in
both frames. Account was also taken of changes in magnitude due to the
overlaying of artificial stars onto natural stars; the completeness
fraction in the brighter magnitude bin was increased accordingly. To
reduce the Poisson uncertainties in the derived completeness
fractions, between 5 and 10 completeness experiments were conducted in
this manner for each color-magnitude bin. Thus, our derived
completenesses for each chip and each color-magnitude bin derive from
between 1000 and 2000 simulated stars. In total, $\sim 10^5$
artificial stars were added for each cluster.

	The artificial stars were distributed in such a manner that
the local surface density was increased by no more than 5\% over its
natural value. Consequently, the distribution of added stars as a
function of radius resembled the King-like profiles of the underlying
cluster surface density distributions, with most of the stars being
added to the central regions where incompleteness is most severe.
Artificial stars were added eight at a time to each annular bin,
equally spaced in position angle but with random phase differences
(extending to sub-pixel scales) between successive radial bins and
from one simulation to the next.  The completeness fraction as a
function of radius and magnitude at a fixed color for the PC frame of
K219 is shown in Figure \ref{fig:completeness}.

	Star counts were carried out in logarithmically-spaced,
annular bins as shown in Figure \ref{fig:layout}. For each star
counted, the completeness appropriate to its radius, magnitude, and
color was estimated using the nearest completeness point in radius,
and bilinear interpolation of the four nearest points in the
color-magnitude grid. Each natural star of suitable color was then
divided by the computed completeness fraction to obtain the actual
surface density of stars in each annular bin. The positions of the
detected stars and the areas encompassed by different annuli were
computed using the distortion coefficients of Holtzman \etal (1995a).
The counts were limited to stars with $V < 25.5$, this limit being
chosen to give a reasonable degree of overlap with the aperture
photometry results without requiring excessively large ({\it i.e.}
more than a factor of five) completeness corrections.  Roughly
speaking, $V = 25.5$ corresponds to a completeness fraction of
$\approx50\%$ for red stars at a radius of $3\arcsec$.

\subsection{Surface Density of Field Stars}

	Contamination of the star counts by field stars (the majority
of which reside in the halo of M31) was not straightforward to
estimate for three reasons: ($i$) the clusters were not centered in the
PC field of view, ($ii$) the tidal radii of the sample clusters are
similar in extent to the size of the PC frame, and ($iii$) some
clusters are close enough to M31 that the spatial distribution of
field stars is significantly nonuniform.

To determine the gradient in the surface density of field
stars, we used DAOPHOT II to find and carry out aperture photometry of
all objects in the WF frames. We found that, over the field of view of
WFPC-2, the distribution of field stars does not depart
significantly from a linear function of the x and y pixel
position. Consequently, after applying the appropriate completeness
corrections, we modeled the surface density $z$ of field stars with a
plane of the form

\begin{equation}
z = c_0 + c_1 x + c_2 y.
\end{equation}

\noindent where the ratio $c_1/c_2$ was fixed by specifying the
direction of the maximum gradient in the surface density of field
stars. This direction was independently determined by using the
Digitized Sky Survey to examine M31 isophotes local to each cluster.
In addition to reducing the number of free parameters, defining the
orientation of this plane in advance allowed us to more easily
identify regions with anomalously high or low surface densities.

	After masking regions obviously contaminated by galaxies, star
clusters, or other localized enhancements, Equation 1 was individually
fitted to the distributions in each of the three WF frames. The
coefficients were then ($i$) averaged, ($ii$) scaled to match the
pixel-scale of the PC frame, and ($iii$) normalized  to the 
surface density of stars measured in a $9\arcsec \times 33\arcsec$ box
on the side of the PC furthest removed from the center of the
cluster. The ability to predict the density of field stars at
arbitrary locations in the field enabled us to carry out star counts
in annuli which were not complete due to the limited field-of-view of
the PC. The raw star counts are given along with the annular areas,
mean annular background level, average completeness corrections, and
corrected surface densities in Tables 6 through 9.

\section{Discussion \label{sec:discussion}}

In Figures \ref{fig:k58} through \ref{fig:k219} we show the surface
density profiles derived from both the star counts and the aperture
photometry for each of the four clusters. The aperture photometry has
been scaled to match the star counts in the region of overlap.  The
error bars plotted for the aperture photometry include a 0.3 DN
uncertainty in the sky-background level, and the uncertainties for the
inner two points have been set to half the difference in DN between
them to account for possible miscentering. The error bars shown for
the star counts take account of both Poisson statistics and the
uncertainties in the completeness corrections. The data are shown only
out to the radius at which the surface density first becomes
consistent (within the computed error bar) with zero. The apparent
agreement between the aperture photometry and the star counts in the
region of overlap suggests that our estimates of the sky background
are not grossly in error.

Shown as solid lines in Figures \ref{fig:k58} through \ref{fig:k219}
are the results of 20 deconvolution iterations using the
Lucy-Richardson (Richardson 1972; Lucy 1974) algorithm. Simulations
have shown that deconvolution of WFPC-2 data to this extent permits
reliable recovery of information down to $\approx0.05\arcsec$. The
deconvolved profiles of both K108 and K219 show a decline (relative to
the ``raw'' profile) in the flux measured for the central pixel. This
is due to the presence of one or more relatively bright red giants
within a few pixels of the center. On the other hand, the inner
deconvolved profile for K105 shows a general steepening and a factor
of two increase in the flux measured in the central pixel. This is in
agreement with the results of Bendinelli \etal (1993), who used
deconvolved, pre-refurbishment Faint Object Camera (FOC) images to determine
that K105 (G105 in their paper) has a power-law cusp at the center and
has therefore likely undergone core collapse. Indeed, over the range
$0.045\arcsec \le r \le 0.23\arcsec$, we find a power-law slope of
$-1.08 \pm 0.1$, somewhat steeper than their estimated value of -0.75,
and strongly suggestive of a past or impending gravothermal
catastrophe.

The long-dashed and short-dashed lines in Figures \ref{fig:k58}
through \ref{fig:k219} are the best-fitting, single-mass King models
before and after 2-dimensional convolution with the PC PSF. King-model
fitting was carried out in two stages; owing to significant coupling
between $r_c$ and $r_t$ and the apparent departures of the
observations from King-like behavior at large radii (see below), we
chose to fit core and tidal radii separately. Firstly,
two-dimensional, single-mass King models were generated and convolved
with the PSF appropriate to the position of the cluster on the PC (\eg
Lauer \etal 1991). The PSF itself was generated using DAOPHOT II to
model the morphologies of the brightest $\sim$30 point sources in each
frame. Aperture photometry, in a manner identical to that used for the
real data, was carried out to produce a one-dimensional surface
density profile, and this profile was then compared with the data to
obtain a value of $\chi^2$.  Using a downhill-simplex method, the core
radius and normalization constant were estimated using aperture
photometry data within $1\arcsec$ of the center.  Secondly, the tidal
radius was estimated using both aperture photometry and star count
data by fixing $r_c$ and searching for the value of $r_t$ which gave
the minimum reduced $\chi^2$. Our best-fit values for $r_c$ and $r_t$
were converted to spatial units assuming a distance to M31 of 770 kpc
(Paper I) and are tabulated and compared with those of previous
investigators in Table 10.

The uncertainties in the core and tidal radii were
estimated using Monte Carlo simulations of the data, based on both the
error bars shown in Figures \ref{fig:k58} through \ref{fig:k219} and a
5\% uncertainty in the adopted sky brightness. Irrespective of
departures from King-like behavior, the simulations yielded 90\%
confidence intervals of $\approx1.2\arcsec$ (4.5pc) in $r_t$ for all
clusters. For K58 and K105, the 90\% confidence intervals on $r_c$
were $\approx0.01\arcsec$ (0.04pc), and for K108 and K219 they were
$\approx0.02\arcsec$ (0.08pc). Of course, these uncertainties apply
only in those cases where it is clear that a King model is
appropriate. The power-law cusp in the deconvolved profile of K105,
for example, is not well fit by the King model which best matches the
raw data, and if there is indeed a constant surface brightness core in
this cluster, it is likely to be considerably overestimated by the
value of $r_c$ given in Table 10. Similarly, the formal fitting
uncertainties on $r_t$ are largely irrelevant in those cases where
there is a clear discrepancy between the form of the model and that of
the data.

The values we obtain for the half-light radii are typical of those
found in Milky Way globular clusters ($<r_h> = 4.2$pc, Djorgovski
1993).  It is reassuring that the HST study of K58 by Fusi Pecci \etal
(1994) using deconvolved Faint Object Camera images yields a measurement
of the core radius which is in reasonable agreement with ours. On the
other hand, the half-light radius found by these authors for K58
(2.1pc) is somewhat smaller than the value of 2.8pc we find
here. We attribute this discrepancy to the very small field of view
($11\arcsec$) of the FOC in its $f/96$ mode. Given the extent of the
surface density profile shown in Figure \ref{fig:k58}, it is quite
probable that they overestimated the sky brightness, leading to an
underestimate of the total light and extent of K58. Their value of
$r_h = 2.4$pc for K105 (compared to our value of 2.9pc) is consistent
with this hypothesis.

	With the exception of K108 (which, in projection, lies closest
to M31 and hence is most severely affected by background
uncertainties), all clusters show apparent departures from the best
fitting King model at radii approaching $r_t$. These departures take
the form of an excess of stars near and beyond the fitted $r_t$, and
occur between 3 and 4 orders of magnitude below the central
surface density. K105 and K219 illustrate this effect most
dramatically. Both the star counts and the aperture photometry show a
tidal turn-down beyond $1\arcsec$. However, rather than following the
model profiles to a steep tidal-cutoff, the data continue downwards in a
comparatively shallow, power-law fashion.

Due to the distribution of uncertainties, the best-fitting values of
$r_t$ are heavily influenced by the inner few data points derived from
the star counts. The single-mass King models cannot be made to fit the
extended portions of the profiles due to the characteristic, upward
inflection of high-concentration, King model profiles.  This is
illustrated in Figure \ref{fig:k219}, where we have plotted a King
model with core radius and normalization identical to that of the King
model which best fits the data for K219, but with the tidal radius
increased to $24\arcsec$ (90pc). Although this model passes through the
outermost points, the reduced $\chi^2$ is very much greater than for
the best-fit case. Clearly, no renormalization of this extended model
would improve the overall fit to the data. 

Note that the cluster isophotes have ellipticities ranging from
$\sim0.07$ (K58) to $\sim0.18$ (K219). However, even if we assumed
that the tidal cutoff radius varies with position angle like the
isophotes, the tidal radius measured along the major axis of the
cluster should differ by no more than 10\% from the tidal radius
predicted by a circular model. The differences between the radius of
the last measured, non-zero star count datum and the radius at the
same surface density of the best-fitting King model exceed what one
might have expected purely from cluster ellipticity by factors
of $\sim8$ (K219) to $\sim15$ (K105). Moreover, K108 shows moderate
isophote ellipticities ($\sim0.12$), but the star counts near $r_t$
are well fit by a King model. Hence we conclude that the
departures near $r_t$ between the models and the star counts are not 
related to the ellipticities seen in the isophotes.

  The departures from King models are consistent, both in magnitude
and in form, with the findings of Grillmair \etal (1995), who studied
a sample of 12 Galactic globular clusters. They concluded that these
extended profiles result from stars which have been stripped from
their parent clusters and are in the process of migrating outwards
along the tidal tails. The power-law profiles seen in their sample
were found to exhibit a variety of slopes, consistent with N-body
models of clusters on eccentric orbits (Grillmair \etal 1996). The
magnitude of the effect and the slope of the power-law profile are
functions of orbital phase and viewing angle. Owing to the much
smaller number of resolved cluster stars in the present work (due to
the much brighter limiting absolute magnitude), the uncertainties in
the counts near $r_t$ are significantly larger than for the sample of
Grillmair \etal (1995). Similarly, the very small numbers of
``extra-tidal'' stars relative to the background preclude a study of
their two-dimensional distribution. However, the systematic behavior
of the profiles shown in Figures \ref{fig:k58} through \ref{fig:k219}
is consistent with the conclusions reached by these authors.

	One might be tempted to dismiss the foregoing on the grounds
that single-mass King models must necessarily be too simplistic to
adequately model the behavior of real clusters. However, we would
argue that multi-mass King models are unlikely to be able to account
for the observed profile shapes near $r_t$ due to the very small range
of stellar masses represented among the counted stars. Likewise, the
aperture photometry is dominated by giant branch stars, so much so
that stars with $V < 26.5$ (the regime sampled by the star counts)
constitute $\approx80\%$ of the total light from each cluster. We note
that $\approx50\%$ of the light is contributed by stars above the
horizontal branch, while only $\approx20\%$ of the
completeness-corrected star counts occur above the same magnitude
level. Nonetheless, the difference in mean stellar mass over this
magnitude interval is entirely inconsequential from the standpoint of
dynamical segregation.

K105 is somewhat analogous to M15 in having both a collapsed core and
a pronounced excess of extra-tidal stars. That both phenomena should
be observed in each of these clusters may not be surprising; the
early onset of core collapse and a rapid rate of repopulation of the
tidally stripped region of a cluster both rely on a relatively short
2-body relaxation time. 

\section{Summary}

	We have analyzed WFPC-2 images of four globular clusters
associated with M31 to obtain surface density profiles and structural
parameters. Using aperture photometry and star count analyses, we
find that:

\begin{itemize}

\item{} there is no evidence for color gradients in the aperture
photometry, 

\item{} for three of the clusters the inner portions of the
1-dimensional surface density profiles can be reasonably-well modeled
using single-mass King models,

\item{} we confirm the discovery by Bendinelli that K105 is likely to
have undergone core collapse, and

\item{} for three of the clusters the surface density profiles depart
from King models at large $r$ in a manner which suggests the presence
of tidal tails.

\end{itemize}

	The detections of a collapsed core and possibly of tidal tails
in extragalactic globular clusters are yet more tributes to the
capabilities of HST. Since proper-motion studies are unlikely to yield
much useful information in the near future for globular clusters in
M31, a deep study of the tidal tails may be the only way to expand our
knowledge of the shapes of the cluster orbits. This in turn would
enable us to infer the distribution of mass in M31's halo, and perhaps
to answer some of the lingering questions concerning the origins of
globular clusters and the formation and evolution of spiral galaxies.

\acknowledgments

This research was conducted by the WF/PC Investigation Definition
Team, supported in part by NASA Grant No. NAS5-1661.

\clearpage

\figcaption{
$V-I$ Color profiles for the four clusters in our sample. The
error bars reflect both the scatter among 8 sectors within each
annulus, and the effect of a 0.3 DN uncertainty in the sky background.
\label{fig:color}}

\figcaption{
Color-magnitude diagram of K219.  The open circles show the
points at which completeness simulations were carried out. The dotted
line shows the colors and magnitudes corresponding to the 50\%
completeness level at a radius of $2.6\arcsec$ from the center of
K219. Similarly, the dashed line indicates the 50\% completeness level
at a radius of $20\arcsec$.
\label{fig:cm}}

\figcaption{
Completeness fractions as a function of both radius and
magnitude, determined from simulations using the PC images of
K219. Shown here are the results for artificial stars of $V-I$ = 1.5.
\label{fig:completeness}}

\figcaption{
Division of the WFPC-2 into annular and partial annular
bins. Though not obvious on this scale, the annuli shown do take into
account the field distortions induced by the camera optics. The gaps
apparent between frames result from having uniformly excluded the
first 60 x and y pixels in each chip so as to remain well clear of the
pyramid shadow.
\label{fig:layout}}

\figcaption{
Surface density as a function of radius computed from the
aperture photometry and star counts for K58. The open circles show the
aperture photometry, scaled to match the star counts in the region of
overlap. The filled circles are completeness-corrected surface
densities derived from the star counts and are shown only out to the
radius at which the background-subtracted surface density first
becomes consistent with zero (to within the uncertainties). The solid
line shows the aperture photometry after 20 Lucy-Richardson
deconvolution iterations. The long-dashed line corresponds to the
best-fitting King model (with parameters taken from Table 10), and the
short-dashed line shows the same model after convolution with the
WFPC-2 PSF.
\label{fig:k58}}

\figcaption{
Same as Figure 5, but for K105.
\label{fig:k105}}

\figcaption{
Same as Figure 5, but for K108.
\label{fig:k108}}

\figcaption{ 
Surface density as a function of radius computed using
aperture photometry and star counts in K219. Symbols are as in Figure
5. In addition, the dotted line shows a King model with the same core
radius and normalization as the solid line, but with a tidal radius
increased to $24\arcsec$ so as to pass through the outermost, non-zero
data points.
\label{fig:k219}}












\end{document}